# MRI-compatible electromagnetic servomotors for image-guided robotic procedures

Lorne W. Hofstetter,[1]* Rock Hadley,[1] Robb Merrill,[1] Huy Pham,[1] Gabriel C. Fine,[1] Dennis L. Parker[1]

[1]Department of Radiology and Imaging Sciences, University of Utah, Salt Lake City, Utah, USA
*email: lorne.hofstetter@hsc.utah.edu

## ABSTRACT
Combining the unmatched soft-tissue imaging capabilities of magnetic resonance imaging (MRI) with high precision robotics has the potential to improve the accuracy, precision, and safety of a wide range of image-guided medical procedures. However, the goal of highly functional MRI-compatible robotic systems has not yet been realized because conventional electromagnetic servomotors used by medical robots can become dangerous projectiles near the strong magnetic field of an MRI scanner. Here we report a novel electromagnetic servomotor design that is constructed from non-magnetic components and can operate within the patient area of clinical scanners. We show that this design enables high-torque and precisely controlled rotary actuation during imaging. Using this servomotor design, an MRI-compatible robot was constructed and tested. The robot demonstrated that the linear forces required to manipulate large diameter surgical instruments in tissues could be achieved during simultaneous imaging with MRI. This work presents the first fully functional electromagnetic servomotor that can be safely operated (while imaging) in the patient area of a 3 Tesla clinical MRI scanner.

## INTRODUCTION
Magnetic resonance imaging (MRI) can volumetrically image the human body in a non-invasive manner without the use of ionizing radiation (1). The ability to visualize anatomical structure and pathology of soft tissues in exquisite detail, as well provide functional information, have made MRI indispensable for the preoperative planning of neurosurgeries (2–5), orthopedic procedures (6,7), tissue biopsies (8–10), and cancer therapies (11–14). However, preoperatively acquired images can quickly become useless due to procedure-induced changes in the tissue geometry or environment. The introduction of needles, resection of tissues, or performance of a craniotomy to gain surgical access to the brain can result in shifting and deformation of soft tissues in the area of interest (15–19). This tissue shift is particularly problematic for procedures where targeting accuracy is paramount to achieving a favorable outcome (5,20) or when intraprocedural discrimination between diseased and healthy tissue relies on advanced imaging techniques such as MRI (14,18,21).

The development of intraoperative MRI emerged to address limitations associated with using static preoperative imaging for surgical guidance. In 1994, an open 0.5 Tesla (T) MRI design was introduced that allowed direct surgical access to the patient during imaging (22,23). The benefits of this surgical approach quickly became apparent in the resection of glioma brain tumors where maximally resecting the tumor while preserving eloquent brain regions was shown to improve survival (18,24,25). More recently, the improved image resolution and widespread availability of closed-bore and high-field scanners (1.5 T and 3 T) has driven their use for intraoperative MRI (21). However, the closed-bore nature of these systems (60-70 cm bore diameter) limits surgical access to the patient during imaging. Freehand approaches are possible but are ergonomically difficult and can require the physician to reach up to 1 meter into the scanner bore for access. As a work around, patient transport to the imaging region of the MRI or operating rooms equipped with a mobile MRI system (26) are used intraoperatively to confirm critical steps during a variety of procedures. However, this paradigm of move-to-image is reactionary and does not enable concurrent intraoperative imaging for real-time guidance.

To compensate for limited patient access in closed-bore MRI scanners, medical robotic systems have been developed that can operate safely in the scanner bore (16,27–33). The aim of such systems is to combine the precision of robotic-assisted procedures with the clinical benefit of high-resolution intraoperative MRI. However, design of these medical systems is complicated by the strong magnetic field generated by the superconducting magnet of the MRI system. Traditional electromagnetic servomotor actuators that have been refined and vetted over decades of use in industrial automation and commercial medical robots are inherently incompatible with MRI. Ferromagnetic and magnetic material used by conventional electromagnetic actuators can become dangerous projectiles if brought near the magnetic field of the MRI scanner. Hence, to date, medical robots that can operate in the MRI have relied on non-magnetic pneumatic and piezoelectric actuator technologies. However, the limited accuracy of pneumatically controlled actuators that utilize long transmission lines and the potential for significant oscillation and overshoot (27,34) make their use unsuitable where high precision is paramount. The electromagnetic noise generated by the operation of commercially available piezoelectric actuators can interfere with the sensitive receiver hardware of the MRI. These actuators when operating simultaneous with MRI have been shown to reduce the image signal to noise ratio (SNR) by 26-80% (35–37). While specially designed controllers have been used to keep this SNR degradation to below 15% (27), achieving dynamic and smooth proportional actuation via closed-loop control of piezoelectric actuators is not trivial. The inability to use the electromagnetic actuation principles that are mainstays of industrial automation has limited the development, functionality, and adoption of medical systems that combine the benefits of robotic precision with the capabilities enabled by high-resolution intraoperative MRI.

In this study we present a servomotor that is constructed from non-magnetic materials that is able to unlock the paradigm of utilizing electromagnetic actuation in close proximity to the magnetic field of the MRI system. We show that this actuator design can be operated simultaneously with MRI without degrading image quality and that an optical rotary encoder and servomotor controller enable closed-loop control. We then demonstrate in a proof-of-concept MRI-compatible surgical robot that this electromagnetic servomotor actuator can be used to drive a biopsy introducer to a target of interest while imaging at 5 frames/second.
These results constitute an important step towards highly functional robotic systems that can be used to perform interventional procedures under concurrent intraoperative MRI guidance.

## RESULTS
**MRI-compatible Direct Current Motor Concept**
The conventional direct current (DC) electromagnetic motor (Figure 1A) is comprised of magnetic and ferromagnetic materials which can become hazardous projectiles if brought near the superconducting magnetic field of an MRI scanner. While many of the magnetic components can be replaced by non-magnetic counterparts, two serve important electromagnetic functions. Permanent magnets inside the motor housing (Figure 1A) produce a static magnetic field that interacts with the electrical current in the rotor windings to generate rotary actuation. The rotor (Figure 1A) is made from ferromagnetic laminations which focus the magnetic flux and thereby enhances the torque generation between rotor windings and the permanent magnets. While these two magnetic motor components serve both necessary and useful functions, their use near the patient area of MRI systems is inherently incompatible due to the strong forces exerted on the components by the field of the MRI system.

Figure 1B presents an electromagnetic motor concept that can operate within the bore of MRI systems and does not require the use of permanent magnets or a ferromagnetic rotor. The motor is designed to utilize the magnetic field generated by the superconducting magnet of the MRI scanner which obviates the need for permanent magnets. Since the strong magnetic field (1.5 T – 3 T for standard clinical systems) is homogeneous within the bore and extends well beyond the patient area, the use of a ferromagnetic rotor to focus magnetic flux is not needed to maintain a strong magnetic flux density at the rotor windings. Removing the ferromagnetic rotor has additional benefits in that it eliminates unwanted motor cogging torque that is associated with the reluctance of ferromagnetic materials (38). Reduction in motor cogging can be particularly important for robotic applications requiring high precision.

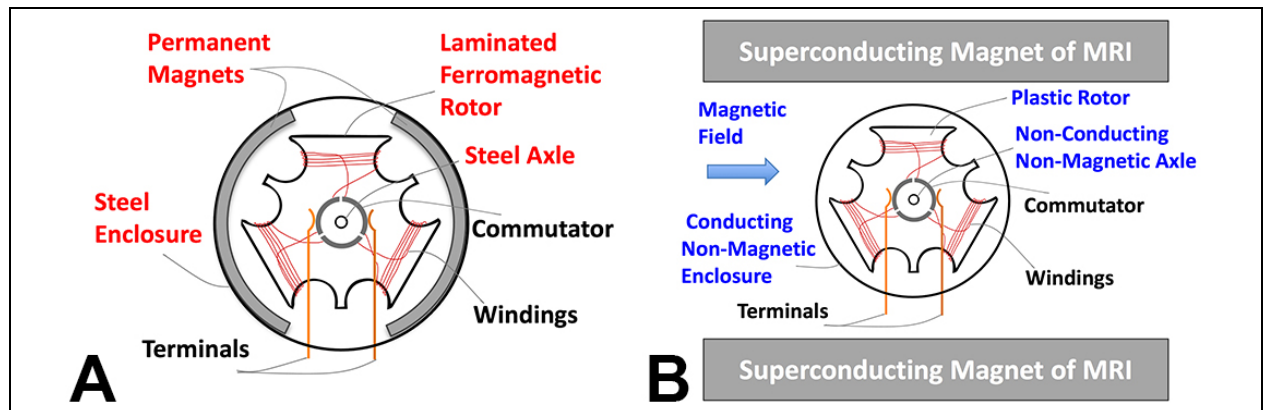

**Fig. 1. Conventional and MRI-compatible electromagnetic direct current motors. (A)** Schematic of conventional direct current (DC) motor. Red labels denote components that are inherently incompatible near strong magnetic fields. **(B)** Schematic of MRI-compatible DC motor concept. Static field of main superconducting magnet is used instead of permanent magnets for torque generation. All magnetic and ferromagnetic materials are replaced by non-magnetic counterparts (blue text).

**MRI-compatible Electromagnetic Servomotor Design and Performance**

We combined the electromagnetic motor concept in Figure 1B with a non-magnetic optical encoder and motor controller (schematic in Figure 2A) to achieve closed loop servomotor functionality in the MRI. Figure S1 shows the servomotor components prior to assembly and the assembled servomotor is shown in Figure 2B with an end view of the servomotor encoder shown in Figure 2C. The encoder consists of two transmissive optical sensor with phototransistor output (TOSwPO) units that detect changes in position and direction of the 3-leaf encoder disk attached to the motor axle. The encoder subdivides each axle revolution into 12 increments. The encoder circuit used to detect changes in the TOSwPO sensors is in the upper right of Figure 2D and the corresponding circuit diagram is shown in Figure S2. Control of rotor position is achieved using a proportional integral (PI) controller implemented on the Arduino microcontroller (shown in Figure 4D) which receives inputs from the encoder circuit. Speed and directional control of the motor is achieved using an H-bridge controller (shown in Figure 4D) which sends a pulse-width modulation (PWM) control signal to the servomotor. A shielded Cat7 ethernet cable electrically connects the servomotor to the controller unit. The final motor control and servomotor assembly including all electromagnetic interference (EMI) shielding and radio frequency (RF) cable traps is shown in Figure 2E.

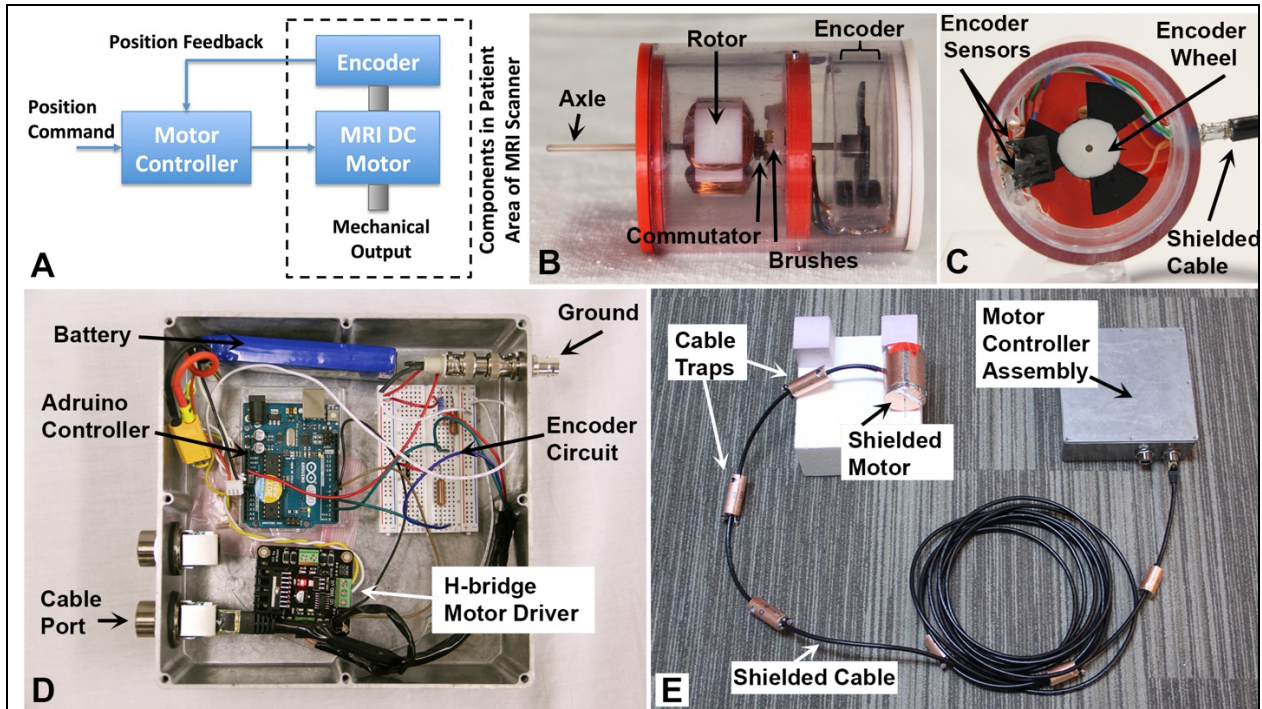

**Fig. 2. MRI-compatible servomotor.** (**A**) Block diagram of the MRI-compatible electromagnetic servomotor concept that enables closed-loop rotary actuation. (**B**) Photograph of servomotor without EMI shielding. Servomotor includes the MRI-compatible DC motor described in Figure 1B and an MRI-compatible encoder. (**C**) End view of the servomotor showing the two TOSwPO encoder sensor units and an encoder disk divided into 60° increments. (**D**) Photograph of motor controller assembly and associated sub-components. (**E**) Photograph of servomotor unit connected to motor controller assembly. All EMI shielding including faraday cage around servomotor and cable traps are used in this final configuration.

Connecting a 7.4 Volt (V) lithium polymer (LiPo) battery (Hobbyking, Hong Kong) to the DC motor terminals induces rotary motion of the motor rotor and axle (Video S1). Video S2 shows that motor operation can be achieved at different orientation angles in the MRI bore. Precise control of the servomotor using closed-loop-feedback control is demonstrated in Video S3. After a command was issued by the motor controller to increment by 120 steps, return to the desired setpoint was maintained even when the motor shaft was forcefully perturbed away from the setpoint. These results demonstrate precise control of an electromagnetic servomotor constructed completely from non-magnetic components and operated inside the MRI scanner bore.

Servomotor performance metrics when powered by the 7.4 V battery and operated at field strength of 2.89 T are shown in Table 1. Stall torque (measured using apparatus shown in Figure S3) and unloaded shaft speed are sufficient for many actuation applications. Servomotor diameter and length are 58, and 74 mm, respectively.

**Simultaneous Imaging and Servomotor Operation**
The MRI transmit/receive hardware is extremely sensitive to radiofrequency (RF) energy. Sources of electromagnetic noise near the proton Larmor frequency (123.23 MHz @ 2.89 T) can significantly degrade imaging performance by introducing unwanted electromagnetic signal into the image receiver hardware. The making and breaking of electrical contacts between the brushes and stator during servomotor operation generates broadband radio frequency (RF) noise over a wide frequency band and this noise source can degrade MRI image quality if not sufficiently mitigated (Figure S4). The H-bridge motor controller uses pulse width modulation to control the effective voltage signal to the motor leads. The square voltage waveforms associated with the PWM control scheme can generate broadband RF noise which is a potential noise source that can contribute to image degradation.

|  | No Load | Stall |
|---|---|---|
| **Current (A)** | 0.07 | 1.58 |
| **Voltage (V)** | 8.29 | 7.65 |
| **Speed (rpm)** | 1524 | - |
| **Torque (mNm)** | - | 73.1 |

**Table. 1. Servomotor performance measurements while operating in MRI-system with a field strength of 2.89 Tesla.**

To minimize interactions between servomotor and the MRI system, three critical design aspects were incorporated into the servomotor and controller shown in Figure 2E. First, EMI shielding principles were used to prevent broadband energy produced by the H-bridge controller and motor brushes from radiating to the MRI receiver hardware. The servomotor was housed in a continuous copper shield (Figure 2E). A 2mm hole in one end of the shield allowed the motor axle to penetrate the housing. The motor controller unit and associated electronics were enclosed in a grounded and shielded box. Power and control signals between motor and motor controller unit were transmitted by a double-shielded Cat7 ethernet cable (4 twisted pairs, one pair supplies current to run the motor, one pair is used to power the encoder diodes, and two pairs are used to return sensor signals to the motor controller). The shield of the Cat7 cable was soldered to the motor shield and electrically connected to the grounded shielded box of the motor controller using RJ45 connectors (Figure 2E). Second, to prevent the motor axle from acting as an antenna and radiating noise contained inside the motor faraday cage, a low conductivity 2mm diameter composite axle was used. Third, six cable traps tuned to the proton Larmor frequency were installed and spaced 15 cm apart on both ends of the shielded cable to prevent any RF energy near 123.23 MHz from traveling on the cable shield. These cable traps serve two important functions: (1) to prevent unwanted common mode currents at the Larmor Frequency from introducing unwanted electrical noise into the imaging region of the MRI system and (2) to minimize potential heating of the shielded cable by the MRI transmit field.

Results from Figure 3 demonstrate that use of the EMI design strategies listed above limits unwanted interactions between the MRI system and the operating servomotor. Measured signal to noise ratio (SNR) of images acquired using MRI differed from control by no more than 1.5% for a range of test configurations during the servomotor operation. The motor position was varied between 45 and 15 cm and the motor was powered by both an H-bridge motor controller and a DC voltage supply. Figure 3C shows that for all test conditions and distances, measured image SNR was remarkably similar to imaging in the absence of the servomotor unit. Thus, the ability to simultaneously image with MRI and operate electromagnetic servomotor actuators using conventional actuation principles and motor controllers was demonstrated.

**Proof of Concept Biopsy Introducer Robot**
The accuracy of using focal biopsy techniques to classify cancer risk critically depends on the lesion targeting accuracy of the procedure (39). Current MRI-guided prostate and breast biopsy procedures employ the use of a grid template and pre-treatment images to guide placement of an introducer sheath in order to obtain access to the desired biopsy target. This sheath placement is an iterative process that consists of moving the patient in and out of the scanner bore for imaging and then manually adjusting the introducer sheath position until the desired target is reached. Once the target is reached, the biopsy can subsequently be obtained.

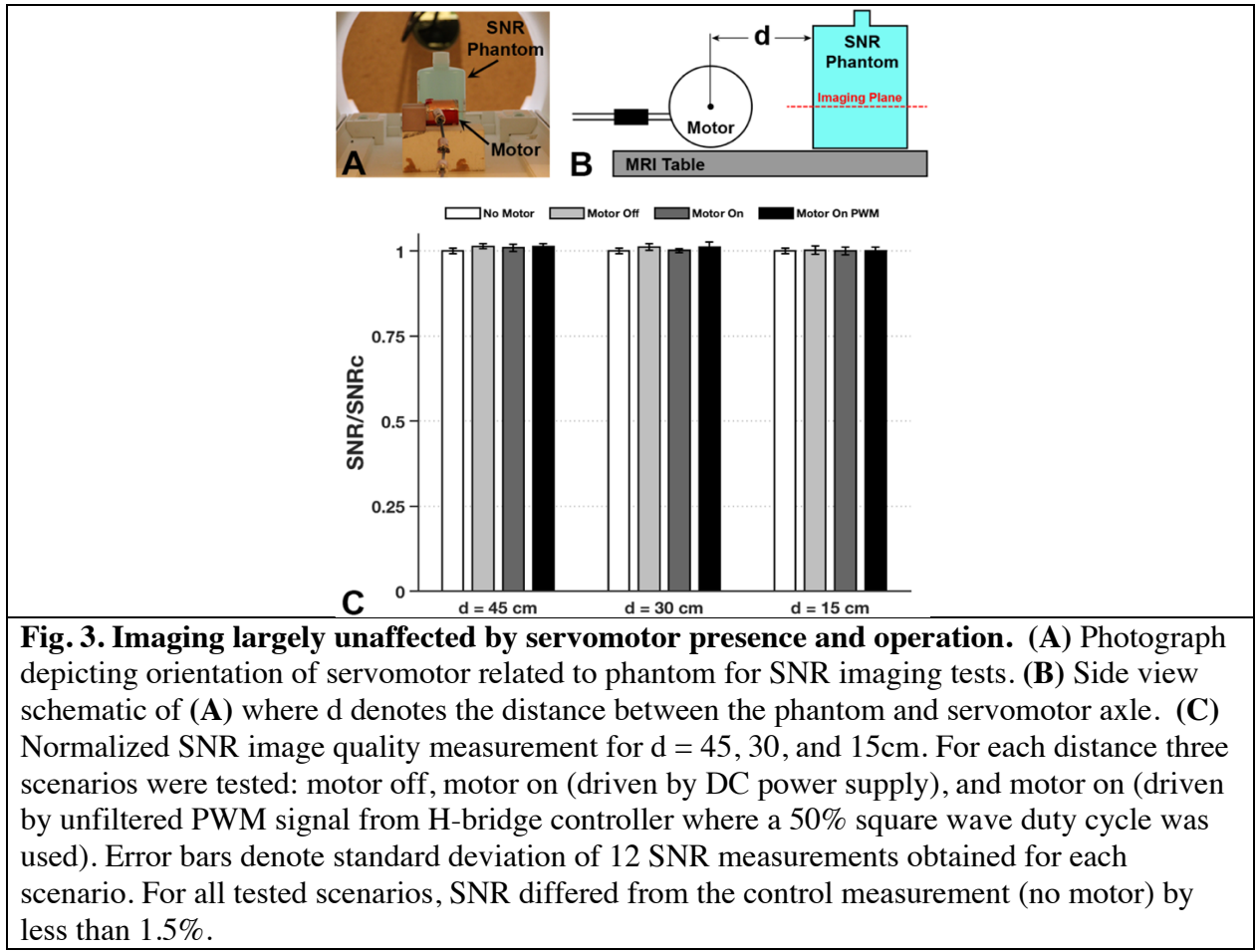

**Fig. 3. Imaging largely unaffected by servomotor presence and operation.** (**A**) Photograph depicting orientation of servomotor related to phantom for SNR imaging tests. (**B**) Side view schematic of (**A**) where d denotes the distance between the phantom and servomotor axle. (**C**) Normalized SNR image quality measurement for d = 45, 30, and 15cm. For each distance three scenarios were tested: motor off, motor on (driven by DC power supply), and motor on (driven by unfiltered PWM signal from H-bridge controller where a 50% square wave duty cycle was used). Error bars denote standard deviation of 12 SNR measurements obtained for each scenario. For all tested scenarios, SNR differed from the control measurement (no motor) by less than 1.5%.

To demonstrate the ability of our MRI-compatible electromagnetic actuator to control a surgical tool during imaging, a proof-of-concept surgical robot was constructed that can place a 9-gauge biopsy introducer sheath under real-time MRI-guidance. An illustration of the 1-degree-of-freedom robot is shown in Figure 4A-B where Figure 4A shows the introducer in a retracted position, and Figure 4B shows the introducer in the fully inserted position. The constructed robot is shown in Figure 4C. The Vernier scale on the introducer stage (shown in Figure 4D) is used to calibrate the position of the linear stage controlling the introducer placement. Gearing connecting the servomotor output to the linear stage results in a maximum linear stage speed of 10 mm/s and a maximum insertion force of 585 N (131 lbs). The maximum range of the linear stage travel is 10 cm.

Simultaneous actuation of the linear stage and imaging with MRI was demonstrated using the setup in Figure 4E. The introducer shown in Figure 4C was replaced by a mock introducer (in Figure 4E) which has an imaging fiducial marker at the position of the mock needle tip. Using a 5 frame/second imaging protocol, imaging was performed during movement of the mock introducer. Images separated by 1 second are shown in Figures 4F-J which demonstrates that high-frame-rate imaging can be used to visualized robot actuation. This simultaneous imaging and actuation are shown in Video S4. Video S5 shows that the orientation of the robot relative to the magnetic field of the MRI can be modified to achieve movement of the introducer along different insertion orientations.

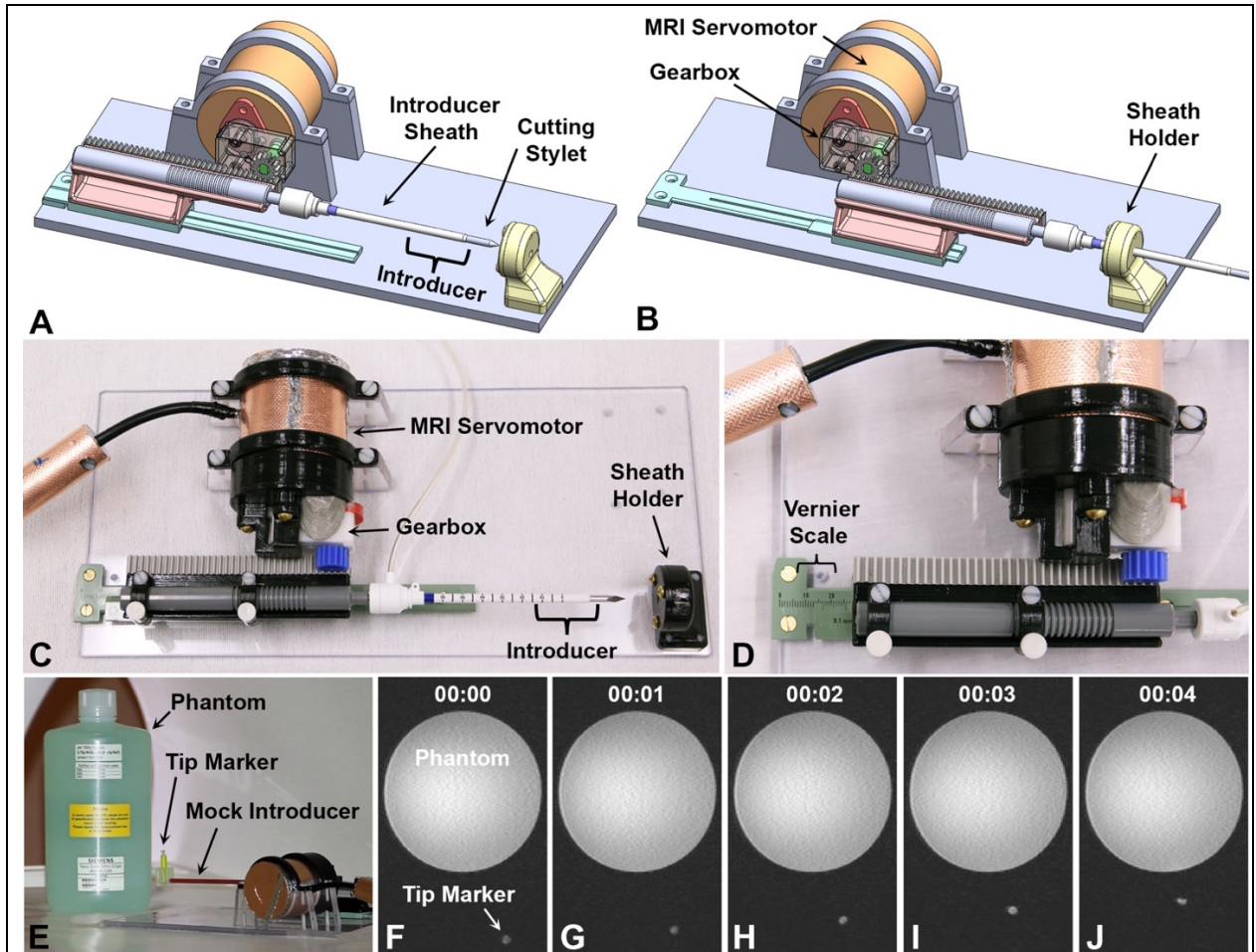

**Fig. 4 Biopsy introducer robot actuated by MRI-compatible electromagnetic servomotor.**
**(A)** Schematic of single degree-of-freedom biopsy introducer robot prior to introducer insertion. **(B)** Schematic showing maximum insertion depth. Sheath holder allows introducer sheath position to be maintained during removal of cutting stylet. **(C)** Photograph of biopsy introducer robot with MRI-compatible servomotor. **(D)** Zoomed in photograph showing Vernier scale on a linear slide with 0.1mm increments. This Vernier scale is used for system calibration when the robot is first powered on. **(E)** Photograph of biopsy robot in the MRI scanner where introducer was replaced with a mock introducer and a tip maker that can be imaged with MRI. Servomotor was used to advance the mock introducer in a controlled manner during continuous imaging at 5 frames per second. Simultaneously acquired coronal images showing movement of tip marker are shown at 1 second increments during actuation in **(F-J)**.

The MRI-compatible biopsy insertion robot was then used to place a 9-gauge introducer sheath to a pre-determined tissue target during continuous imaging. Volumetric MRI was performed prior to needle insertion to determine the desired introducer sheath placement location in imaging coordinates. The robot was commanded under one continuous operation to drive the cutting stylet and introducer sheath from the initial position (Figure 5A) to the desired target (Figure 5B) and then to remove the cutting stylet from the introducer sheath (Figure 5C). The corresponding real-time images for each of these steps are shown in Figure 5D-F with the pre and post introducer sheath placement images shown in Figure 5G and 5H, respectively. This ex vivo tissue experiment demonstrates that a proof-of-concept surgical robot powered by the MRI-compatible electromagnetic servomotor can drive a large diameter introducer through tissue to reach a desired and predetermined target. The full sequence for the introducer sheath placement with simultaneous MRI is shown in Video S6.

Page **7** of **20**

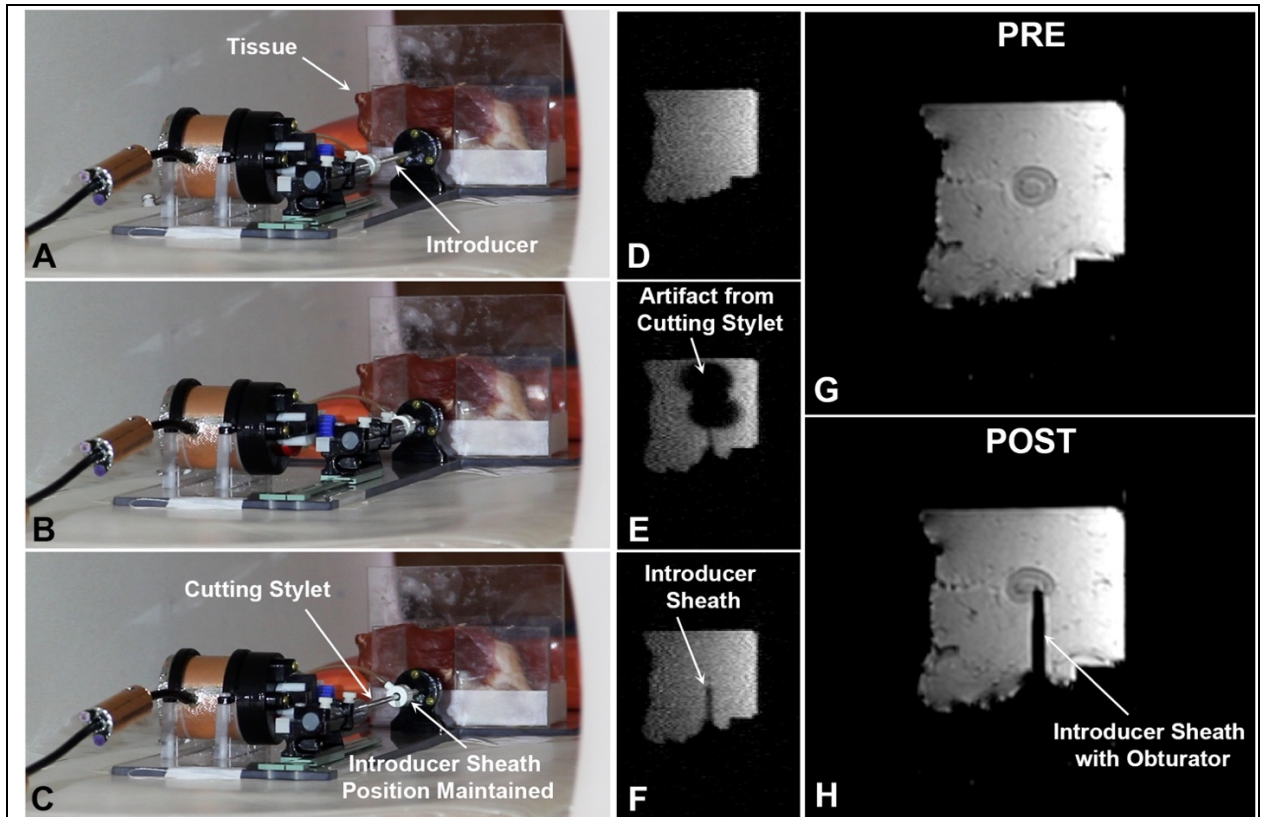

**Fig. 5. Biopsy introducer robot with MRI-compatible servomotor places introducer sheath at desired target location in ex vivo tissue.** Photographs of introducer robot prior to introducer insertion **(A)**, at target insertion depth **(B)**, and during cutting stylet removal **(C)** are shown. The corresponding simultaneous MR images obtained during robot operation are shown in **(D)**, **(E)**, and **(F)**, respectively. **(G)** Sagittal MR image of tissue sample showing target lesion prior to biopsy introducer placement. **(H)** Sagittal MR image following introducer sheath placement showing that desired placement position of introducer sheath was achieved.

## DISCUSSION

We have presented an electromagnetic servomotor that can safely operate in the patient area of MRI scanners. The servomotor was constructed from non-magnetic materials and hence is not a potential projectile hazard in this environment. Rotary actuation was generated by leveraging the interaction between electrical currents in the servomotor rotor windings and the superconducting magnetic field of the MRI scanner. Closed-loop position control of the servomotor axle was achieved using an optical encoding method. The servomotor was specifically designed to minimize EMI so that simultaneous robotic actuation and imaging with MRI could be performed. Utilizing this servomotor, a proof-of-concept robot was constructed and tested to show that the linear forces required to manipulate large diameter surgical instruments in tissues could be achieved during simultaneous serial imaging with MRI.

A key benefit of this MRI-compatible actuator technology is that it uses standard electromagnetic actuation principles and control hardware commonly used in commercial medical robots and industrial automation. Thus, the ability to draw on this prior body of work will simplify the future development of highly functional MRI-compatible robotic systems.

As a second major benefit, the servomotor enables simultaneous imaging and robotic actuation. Under the current paradigm of high-field intraoperative MRI, either the patient or MRI scanner is moved into position for imaging to confirm critical steps in the procedure. This repositioning is both time consuming and reactionary in that it does not enable real-time decision making in an ever-changing surgical environment. For example, brain shift during neurosurgery occurs continuously and unpredictably throughout the procedure (40,41), and for procedures that can benefit from the visualization capabilities of interoperative MRI, such as glioma tumor resections and needle-based procedures, robotic systems that enable serial and concurrent imaging could improve procedural precision and safety. MRI-guided prostate biopsy studies have shown that needle bending and skin-needle interactions during manual insertion of the biopsy needle can result in targeting errors that are on the order of half a centimeter (17,42). A robotic-assisted needle insertion approach that utilizes feedback from concurrent intraoperative MRI could be used to visualize and correct for needle bending and tissue deformation in real-time during needle insertion. This would minimize the need for needle reinsertion following a failed targeting attempt. The MRI-compatible servomotor presented in this work may enable the use 4D intraoperative MRI where volumetric MRI is performed serially in time and is used to inform control of robotic systems during medical procedures.

Leveraging the strong magnetic field of the MRI system to actuate servomotors enables safe electromagnetic actuation in the patient area of the scanner. The 1.5 or 3 Tesla magnetic flux density produced by conventional close-bore MRI systems unlocks the possibility of small footprint servomotor actuators. Current state of the art electromagnetic motors use rare-earth metals such as Neodymium to generate the magnetic field that interacts with electrical currents in the motor windings. The remanence of a Neodymium magnet is less than 1.3 Tesla (43) and the resulting magnetic flux density in air rapidly decreases with distance from the magnet surface. In a servomotor design that leverages the superconducting field of the MRI, the in-air magnetic flux density that is produced is much larger than the in-air magnetic field that can be generated by the rare-earth permanent magnets. Hence, an MRI-compatible servomotor as described can in principle be constructed with a much smaller physical footprint while maintaining the same peak operating performance of state-of-the-art permanent magnet servomotors. This potential for miniaturization is important for developing robotics system that can operate in the limited space of closed-bore MRI systems.

Using the strong magnetic field of the MRI for actuation does place some constraints on the orientation of the servomotor during operation. The MRI-compatible servomotor presented here used a mechanical commutation scheme which requires that the rotational alignment of the motor brushes and static magnetic field of the MRI be maintained for optimal motor performance. However, for applications where the rotational alignment of the servomotor cannot be maintained, a redesign of the commutator would allow consistent operation for any rotational alignment. Instead of mechanical commutation, the use of slip rings to provide continuous electrical connections to the rotor, combined with the electrical commutation control schemes widely used by brushless permanent magnet motors would eliminate the need to maintain a strict commutator orientation.

Other translations of the servomotor that change the amount of flux passing through the rotor loop windings will alter the achievable torque and speed of the motor. If the servomotor coils are powered by a constant voltage source, a change in the motor orientation that decreases the magnetic flux through the rotor loops will decrease the maximum stall torque and increase the unloaded motor speed. However, if the servomotor is sufficiently sized to accommodate the different orientations needed for a robotic application, this change in operating conditions with orientation can be readily controlled by the closed-loop nature of the servomotor controller. This is demonstrated by Videos S4 and S5 where orientation of the proof-of-concept robot was changed with respect to the superconducting field of the MRI system. Videos S2 and S3 demonstrate that changes in servomotor orientation that do not change the rotational alignment of the brushes or alter the magnet flux seen by the rotor windings have no impact on the motor operation. Given this flexibility and the possibility of using electrical commutation schemes that enable an arbitrary rotational alignment of the servomotor, a very wide range of actuation options at different orientations are possible.

We also anticipate that this MRI-compatible servomotor will have important non-surgical applications. There are, for example, many non-surgical applications that require motion in the MRI scanner. These include magnetic resonance elastography (44), which requires a vibrating pillow next to the patient, MRI compatible ultrasound devices for imaging and therapy that could be positioned robotically, and phantom studies that require internal motions to mimic physiological motion.

# MATERIALS AND METHODS

The objective of this study was to demonstrate that electromagnetic actuator principles vetted and widely used in industrial automation are not inherently incompatible with MRI systems. This study design was to: (1) Build an electromagnetic servomotor that uses the field generated by the superconducting magnet of a clinical MRI system for actuation to achieve controlled rotary motion; (2) Measure the torque and speed characteristics of the servomotor while operating in the patient area of an MRI scanner; (3) Implement EMI reduction and shielding strategies to enable simultaneous operation of servomotor and MRI; (4) Quantify interactions between the operating servomotor and MRI using SNR measurements; (5) Construct a proof of concept biopsy introducer surgical robot using the presented electromagnetic servomotor design and demonstrate that real-time MR imaging can be used to track robot motion; and (6) Demonstrate that a biopsy introducer robot can drive and place a 9-gauge introducer sheath to a desired target location in an ex vivo tissue sample. For all experiments in this study, the servomotor and robot were operated while in the patient area of a 60 cm diameter bore, clinical 3T Prisma Fit MRI scanner (Siemens Medical Solutions, Erlangen, Germany).

**Servomotor Construction Details**
Components of the prototype servomotor are shown in Figure S1. The motor axle was constructed from a 2 mm diameter G-10/FR4 non-conducting rod (McMaster-Carr, #8669K627). Mechanical commutator and brushes were obtained from a disassembled 280 micro 3V-12V DC toy motor. The support structure for the rotor windings was 3D printed from VeroWhitePlus (Stratasys, Israel). Each of three 100-turn rotor windings (~20 mm$^2$ cross-sectional area) was hand wound from 30-gauge Polyamideimide magnet wire (Remington Industries, Illinois, USA). Once wound, cyanoacrylate glue was used to secure rotor windings in place. Solder was used to connect rotor windings to the commutator. The measured resistance of each rotor loop was 1.2Ω.

The outer housing of the servomotor was constructed from 57.1 mm outer diameter clear polycarbonate tubing (McMaster-Carr, #8585K28). Motor end rings were 3D printed from ABS plastic and 2 mm inner diameter Olite bushings (McMaster-Carr, #6658K411) were pressed into the motor end rings to provided support for the motor axle. Powdered graphite lubricant (Panef Corp., Milwaukee, WI) was used to minimize friction between the axle and bushings. One end ring had 3D printed details to enable proper alignment and fixation of the brushes to the end ring. A 50.8 mm outer diameter clear polycarbonate tubing (McMaster-Carr, #8585K26) that fits inside the outer housing was used to maintain alignment and proper separation of motor end rings. Two additional tight-fitting bushing (not shown in Figure S1) were secured to the axle to keep the rotor properly situated between the two end rings of the motor housing.

The encoder was constructed from a 3D printed ABS plastic encoder wheel (5 mm thickness, 35 mm maximum diameter). To ensure opacity of the encoder wheel, each leaf was spray coated with black paint. The two TOSwPO sensors (TCST2103, Vishay Intertechnology Inc.) were mounted to the circular support polycarbonate tubing (50.8 mm outer diameter) so that four unique states of the sensors were possible: (1) both sensors blocked by an opaque encoder leaf, (2) first sensor blocked by an opaque encoder leaf and second not blocked, (3) second blocked by an opaque encoder leaf and first not blocked, (4) neither sensor blocked. These four unique states enabled rotor motion and direction to be measured.

The constructed servomotor assembly without EMI shielding is shown in Figure 2B and 2C. To install the EMI shielding, the outer polycarbonate housing of the servomotor was coated in copper shielding foil tape (3M, #1739-17). Tape seams were soldered to ensure electrical connection between all segments. A 25-foot double-shield Cat7 ethernet cable consisting of 4-twisted pair 26-gauge wires (Tera Grand, California, USA) was used to transmit all signals between the servomotor (Figure 2B) and motor control assembly (Figure 2D). One twisted pair was used to supply current to the motor terminals. A second twisted pair was used to supply power to the diodes on the TOSwPO sensors. The remaining two twisted pairs communicate the TOSwPO sensor signals to the motor controller assembly. All electrical connections at the servomotor were soldered and the Cat7 cable shield was soldered to copper tape on the motor housing. An RJ45 connector on the end of the Cat7 cable distant from the servomotor enabled easy connection of wires and cable shield to the motor controller assembly.

The motor controller assembly (Figure 1D) was housed in an 18.8x18.8x6.7 cm aluminum box. A female RJ45 connecter (PEI-genesis, PA, USA) enabled easy connection of the servomotor to the motor controller assembly. The shielded box, cable shield, and motor shield are all connected to ground via a BNC connector on the back of the box (Figure 2D). All power to the servomotor and controller is provided by a 7.4V 2-cell LiPo 2100 milli-amp-hour (mAh) battery (Hobbyking, Hong Kong). All control logic was implemented on the Arduino Uno Rev3. The TOSwPO sensor signals are fed as inputs to the Arduino which then sends the desired PWM control signal to a 2-amp H-bridge motor controller (DFRobot, DRI00002). Additional circuitry (see circuit diagram in Figure S2) used to read TOSwPO signals is located on a solderless breadboard (see Figure 2D).

Six floating shield current suppression traps (45) were constructed and installed 15-cm apart (Figure 1E) on the terminal ends of the Cat7 ethernet cable in order to suppress common mode currents on the cable shield. RF traps were constructed to attenuate signals at 123.23 MHz (the proton Larmor frequency at 2.89 Tesla) to further minimize interactions between MRI transmit/receive hardware and servomotor hardware. Outer and inner diameter of traps were 22 mm and 7 mm, respectively. Two trap variants with a length of 38 mm and 57 mm had a mean attenuation at 123.23 MHz of 7.4 dB and 11.3 dB, respectively.

**Encoder Sensing and Controller**
Accurate detection of the signal from the TOSwPO sensors is required for precise servomotor control. The schematic in Figure S2 shows the circuit used to power the TOSwPO diodes as well as the voltage dividing sensing circuit. The voltage at the Arduino input pin for a given sensor is low when the opaque encoder wheel leaf is not blocking the transistor sensor, otherwise the voltage is high. Control software was implemented on the Arduino to continuously monitor the state of the sensor signal to allow changes in rotor position to be detected and the number of rotation increments to be counted. The Arduino was programmed to receive a command input (in number of encoder steps) from the user. A proportional integral (PI) controller was then used to generate a control signal sent to the H-bridge motor controller to achieve the desired commanded actuation.

**Servomotor Performance Measurements**
Stall torque and unloaded motor speed was measured for the servomotor operating in the 2.89 T magnetic field of a clinical MRI scanner. For both measurements, the motor was directly powered by the 7.4V LiPo battery. A Fluke 77 and Fluke 27 multimeter were used to measure voltage across the motor leads and rotor current during operation. Stall torque was measured using the experimental setup shown in Figure S3. A mass was attached to the 2-inch diameter pulley mounted on the servomotor axle. The amount of mass was increased incrementally until the maximum lifting capacity of the motor-pulley assembly was determined. The reported stall torque is the product of the maximum lifted weight times the pulley radius. To determine the maximum motor shaft speed, the servomotor was powered in the unloaded state. The encoder hardware and associated circuitry was used to count the number of full axle revolutions occurring over a one-minute interval.

**SNR Measurement during motor operation**
Interactions between the operating servomotor and MRI were evaluated using SNR measurements. The orientation of the motor and phantom being imaged are depicted in Figure 3B. Simultaneous imaging and motor operation was tested for three different states: (1) Motor off but located in the MRI scanner bore, (2) Motor on and powered by the 7.4V LiPo battery, and (3) Motor on and powered by the PWM signal output from H-bridge motor controller. The distinction between (2) and (3) was that the PWM signal from the motor controller generated a 50% duty cycle square wave voltage signal. For each test condition, image quality of the MRI was evaluated by measuring the SNR for each acquired image. SNR measurements were obtained for three different motor-phantom separation distances (d = 15, 30, 45 cm). For each separation distance, the three motor states (described above) were tested. Control SNR measurements (no motor) were also acquired with motor and motor controller completely removed from the MRI scanner room.

For SNR measurements, a 2 channel transmit/receive body coil was used to acquire a 2D gradient echo (GRE) image in the coronal plane with the following acquisition parameters: echo time/repetition time (TE/TR) = 3.58/200 ms, flip angle = 60°, field of view (FOV) = 22 cm, resolution = 1.72×1.72×5 mm, bandwidth = 260 Hz/pixel, 1 average. Images for each coil were reconstructed from raw k-space data in MATLAB (MathWorks, Natick, MA). SNR maps were formed using the noise-covariance-weighted sum of squares magnitude image reconstruction method (46–48). Noise covariance information was calculated from pixels in the over-scan area of the image that was dominated by noise. For each scenario, 12 SNR maps were acquired. For each map, the mean SNR over the phantom cross-section was calculated and reported as a single SNR image quality metric. The mean and standard deviation of this mean SNR image quality metric was calculated and reported for each tested scenario.

**Biopsy Introducer Robot Construction**
A proof-of-concept one degree of freedom robot was constructed from non-magnetic components (Figure 4A-D). A modified plastic Vernier caliper was used as a linear stage and the 0.1mm scaling was used for initial calibration of the linear stage position when the robot was first powered on. The MRI-compatible electromagnetic servomotor described earlier in this paper was used for actuation. A 120:1 Plastic Gearmotor (Pololu, NV, USA) was mounted to the output axle of the servomotor using a 3D printed gearbox holder made from ABS plastic (shown in Figure 4C). To ensure that the gearmotor was non-magnetic, the ferromagnetic steel axles in the gearbox were replaced by 2 mm diameter 316-stainless steel axles (McMaster Carr, #9298K31). A 15 tooth 15mm diameter plastic gear (McMaster-Carr, 2262N415) was attached to the output of the gearmotor and coupled to a matching linear gear rack (McMaster-Carr, 266N57) to provide actuation of the linear stage. A 3D printed sheath holder (in Figure 4B and 4C) allows the biopsy introducer sheath to be held in a fixed position during removal of the cutting stylet. To demonstrate simultaneous imaging and actuation of the robot, a mock introducer (Figure 4E) consisting of a fiberglass rod with cylindrical fiducial marker (Hologic, Marlborough, MA) was constructed.

An ex vivo tissue experiment was performed to demonstrated accurate placement of a 9-gauge introducer (Hologic, Marlborough, MA) into a pre-specified target during simultaneous imaging with MRI. The introducer, which is comprised of a cutting stylet and introducer sheath used for MRI-guided breast biopsy procedures, was mounted onto the robot linear stage as is shown in Figure 4C. A sheath holder (shown in Figure 4C) that uses a rubber friction mechanism was built to both allow the insertion of the introducer and to hold the introducer sheath at the desired insertion depth during removal of the cutting stylet.

### Ex Vivo Experiment: Tissue Sample Preparation and Robot Calibration

Ex vivo porcine loin was obtained from a local grocery store. To emulate a cancerous tissue lesion, a small incision was made and a pitted olive was embedded in the tissue. The tissue sample was placed in a tissue holder (visible in Figure 5A) that is attached to the robot base. The robot was powered on after being placed in the MRI scanner. The Vernier scale on the linear stage was used to calibrate the initial position of the biopsy introducer when the robot was first powered on.

Next, pretreatment MRI using a 3D VIBE pulse sequence was performed with the following scan parameters: TE/TR = 2.46/7.04 ms, flip angle = 10°, field of view (FOV) = 25.6 cm, resolution = 1×1×1 mm, bandwidth = 890 Hz/pixel, 3 averages. Using the MRI console, the image coordinates of the desired biopsy target were selected from the VIBE images. This coordinate position was used by the servomotor controller to determine the number of motor increments needed to place the introducer sheath at the desired insertion depth. Following automated placement of introducer sheath and removal of cutting stylet, a plastic obturator (Hologic, Marlborough, MA) was inserted into the cutting sheath to provide improved visualization of the sheath tip position with MRI. Imaging using the same pretreatment 3D VIBE imaging protocol was performed to confirm placement location of the introducer sheath.

### Rapid MRI Imaging Protocols used During Robot Actuation

Single slice 2D MRI was used to track the mock introducer tip location during the phantom experiment and to actively monitor the introducer insertion during the ex-vivo experiment. The spine coil array mounted in the patient table was used. Pulse sequence parameters were chosen to achieve an imaging rate of 5 frames/second (0.2 seconds per image). For the phantom experiment, a TRUFI pulse sequence was used with the following scan parameters: TE/TR = 1.94/3.87 ms, flip angle = 45°, matrix = 256×62, resolution = 1.17×1.25×5 mm, bandwidth = 1149 Hz/pixel, partial Fourier in phase-encoding = 5/8, GRAPPA parallel imaging with 22 reference lines and an acceleration factor of 2. For the ex vivo experiment, a FLASH pulse sequence was used with the following scan parameters: TE/TR = 2.24/4.9 ms, flip angle = 8°, matrix = 256×58, resolution = 1.17×1.46×5 mm, bandwidth = 1150 Hz/pixel, partial Fourier in phase-encoding = 6/8, GRAPPA parallel imaging with 24 reference lines and an acceleration factor of 2.

### Statistical Methods

All data in bar plots are shown as mean +/- SD.

# REFERENCES

1. Edelman, R. R. & Warach, S. Magnetic Resonance Imaging. *N. Engl. J. Med.* **328,** 708–716 (1993).
2. Wengenroth, M. *et al.* Diagnostic benefits of presurgical fMRI in patients with brain tumours in the primary sensorimotor cortex. *Eur. Radiol.* **21,** 1517–1525 (2011).
3. Hervey-Jumper, S. L. & Berger, M. S. Maximizing safe resection of low- and high-grade glioma. *J. Neurooncol.* **130,** 269–282 (2016).
4. Elias, W. J. *et al.* A Randomized Trial of Focused Ultrasound Thalamotomy for Essential Tremor. *N. Engl. J. Med.* **375,** 730–739 (2016).
5. Rich, C. W. *et al.* MRI-guided stereotactic laser corpus callosotomy for epilepsy: distinct methods and outcomes. 1–13 (2021). doi:10.3171/2020.7.JNS20498.
6. Goyal, N. & Stulberg, S. D. Evaluating the Precision of Preoperative Planning in Patient Specific Instrumentation: Can a Single MRI Yield Different Preoperative Plans? *J. Arthroplasty* **30,** 1250–1253 (2015).
7. An, V. V. G., Sivakumar, B. S., Phan, K., Levy, Y. D. & Bruce, W. J. M. Accuracy of MRI-based vs. CT-based patient-specific instrumentation in total knee arthroplasty: A meta-analysis. *J. Orthop. Sci.* **22,** 116–120 (2017).
8. Kasivisvanathan, V. *et al.* MRI-Targeted or Standard Biopsy for Prostate-Cancer Diagnosis. *N. Engl. J. Med.* 1767–1777 (2018). doi:10.1056/nejmoa1801993
9. Lehman, C. D. *et al.* Clinical experience with MRI-guided vacuum-assisted breast biopsy. *Am. J. Roentgenol.* **184,** 1782–1787 (2005).
10. Liberman, L., Bracero, N., Morris, E., Thornton, C. & Dershaw, D. D. MRI-guided 9-gauge vacuum-assisted breast biopsy: Initial clinical experience. *Am. J. Roentgenol.* **185,** 183–193 (2005).
11. Thibault, F. *et al.* MRI for Surgical Planning in. *Breast* 1159–1168 (2004).
12. McClure, T. D. *et al.* Use of MR imaging to determine preservation of the neurovascular bundles at robotic-assisted laparoscopic prostatectomy. *Radiology* **262,** 874–883 (2012).
13. Park, B. H. *et al.* Influence of magnetic resonance imaging in the decision to preserve or resect neurovascular bundles at robotic assisted laparoscopic radical prostatectomy. *J. Urol.* **192,** 82–88 (2014).
14. Jacobs, A. H. *et al.* Imaging in neurooncology. *NeuroRx* **2,** 333–347 (2005).
15. Maurer, C. R. *et al.* Investigation of intraoperative brain deformation using a 1.5-T interventional MR system: Preliminary results. *IEEE Trans. Med. Imaging* **17,** 817–825 (1998).
16. Schouten, M. G. *et al.* Evaluation of a robotic technique for transrectal MRI-guided prostate biopsies. *Eur. Radiol.* **22,** 476–483 (2012).
17. Moreira, P. *et al.* Evaluation of robot-assisted MRI-guided prostate biopsy: Needle path analysis during clinical trials. *Phys. Med. Biol.* **63,** (2018).
18. Golub, D. *et al.* Intraoperative MRI versus 5-ALA in high-grade glioma resection: a network meta-analysis. *J. Neurosurg.* **134,** 484–498 (2021).
19. Gerard, I. J. *et al.* Brain shift in neuronavigation of brain tumors: A review. *Med. Image Anal.* **35,** 403–420 (2017).
20. Starr, P. A. *et al.* Interventional MRI–guided deep brain stimulation in pediatric dystonia: first experience with the ClearPoint system. *J Neurosurg Pediatr.* **14,** 400–408 (2014).
21. Mislow, J. M. K., Golby, A. J. & Black, P. M. Origins of Intraoperative MRI. *Magn. Reson. Imaging Clin. N. Am.* **18,** 1–10 (2010).
22. Schenck, J. F. *et al.* Superconducting open-configuration MR imaging system for image-guided therapy. *Radiology* **195,** 805–814 (1995).
23. Black, P. M. L. *et al.* Development and implementation of intraoperative magnetic resonance imaging and its neurosurgical applications. *Neurosurgery* **41,** 831–845 (1997).

24. Claus, E. B. *et al.* Survival rates in patients with low-grade glioma after intraoperative magnetic resonance image guidance. *Cancer* **103,** 1227–1233 (2005).
25. Lacroix, M. *et al.* A multivariate analysis of 416 patients with glioblastoma multiforme: Prognosis, extent of resection, and survival. *J. Neurosurg.* **95,** 190–198 (2001).
26. Sutherland, G. R. *et al.* A mobile high-field magnetic resonance system for neurosurgery. *J. Neurosurg.* **91,** 804–813 (1999).
27. Li, G. *et al.* Robotic System for MRI-Guided Stereotactic Neurosurgery. *IEEE Trans. Biomed. Eng.* **62,** 1077–1088 (2015).
28. Patel, N. A. *et al.* An Integrated Robotic System for MRI-Guided Neuroablation: Preclinical Evaluation. *IEEE Trans. Biomed. Eng.* 1–1 (2020). doi:10.1109/TBME.2020.2974583
29. Song, S. E. *et al.* Development and preliminary evaluation of a motorized needle guide template for MRI-guided targeted prostate biopsy. *IEEE Trans. Biomed. Eng.* **60,** 3019–3027 (2013).
30. Kaiser, W. A., Fischer, H., Vagner, J. & Selig, M. Robotic system for biopsy and therapy of breast lesions in a high-field whole-body magnetic resonance tomography unit. *Invest. Radiol.* **35,** 513–519 (2000).
31. Groenhuis, V., Siepel, F. J., Veltman, J., Van Zandwijk, J. K. & Stramigioli, S. Stormram 4: An MR Safe Robotic System for Breast Biopsy. *Ann. Biomed. Eng.* **46,** 1686–1696 (2018).
32. Chinzei, K., Hata, N., Jolesz, F. A. & Kikinis, R. Surgical assist robot for the active navigation in the intraoperative MRI: Hardware design issues. *IEEE Int. Conf. Intell. Robot. Syst.* **1,** 727–732 (2000).
33. Masamune, K. *et al.* Development of an MRI-compatible needle insertion manipulator for stereotactic neurosurgery. *J. Image Guid. Surg.* **1,** 242–8 (1995).
34. Yang, B., Tan, U. X., McMillan, A. B., Gullapalli, R. & Desai, J. P. Design and control of a 1-DOF MRI-compatible pneumatically actuated robot with long transmission lines. *IEEE/ASME Trans. Mechatronics* **16,** 1040–1048 (2011).
35. Fischer, G. S. *et al.* MRI compatibility of robot actuation techniques--a comparative study. *Med. Image Comput. Comput. Assist. Interv.* **11,** 509–17 (2008).
36. Krieger, A. *et al.* Development and evaluation of an actuated MRI-compatible robotic system for MRI-guided prostate intervention. *IEEE/ASME Trans. Mechatronics* **18,** 273–284 (2013).
37. Elhawary, H. *et al.* in 519–526 (Springer, Berlin, Heidelberg, 2006). doi:10.1007/11866565_64
38. Pillay, P. & Krishnan, R. Application Characteristics of Permanent Magnet Synchronous and Brushless dc Motors for Servo Drives. *IEEE Trans. Ind. Appl.* **27,** 986–996 (1991).
39. Robertson, N. L. *et al.* Prostate cancer risk inflation as a consequence of image-targeted biopsy of the prostate: A computer simulation study. *Eur. Urol.* **65,** 628–634 (2014).
40. Nabavi, A. *et al.* Serial intraoperative magnetic resonance imaging of brain shift. *Neurosurgery* **48,** 787–798 (2001).
41. Warfield, S. K. *et al.* Capturing intraoperative deformations: Research experience at Brigham and Women's hospital. *Med. Image Anal.* **9,** 145–162 (2005).
42. Blumenfeld, P. *et al.* Transperineal prostate biopsy under magnetic resonance image guidance: A needle placement accuracy study. *J. Magn. Reson. Imaging* **26,** 688–694 (2007).
43. Pyrhonen, J., Jokinen, T. & Hrabovcova, V. *Design of Rotating Electric Machines*. (John Wiley & Sons, Ltd, 2014).
44. Muthupillai, R. *et al.* Magnetic resonance elastography by direct visualization of propagating acoustic strain waves. *Science (80-. ).* **269,** 1854–1857 (1995).

45. Seeber, D. A., Jevtic, J. & Menon, A. Floating shield current suppression trap. *Concepts Magn. Reson. Part B Magn. Reson. Eng.* **21,** 26–31 (2004).
46. Roemer, P. B., Edelstein, W. A., Hayes, C. E., Souza, S. P. & Mueller, O. M. The NMR phased array. *Magn. Reson. Med.* **16,** 192–225 (1990).
47. Kellman, P. & McVeigh, E. R. Image reconstruction in SNR units: A general method for SNR measurement. *Magn. Reson. Med.* **54,** 1439–1447 (2005).
48. Keil, B. *et al.* A 64-channel 3T array coil for accelerated brain MRI. *Magn. Reson. Med.* **70,** 248–258 (2013).

## ACKNOWLEDGEMENTS

**Funding:** This work was supported by the National Institutes of Health under Grants F30CA228363, R37CA224141, R01EB028316, and S10D018482 and the Mark H. Huntsman endowed chair.

**Author contributions:** L.W.H. developed and built MRI-compatible servomotor, designed experiments, collected data, and drafted the manuscript. R.H. developed the MRI-compatible servomotor, oversaw development of the EMI interference reduction strategies, and designed experiments. R.M. assisted in the mechanical design and build of the MRI-compatible servomotor and designed mechanical hardware for experiments. H.P. designed ex vivo tissue experiment, G.C.F. designed ex vivo tissue experiment. D.L.P. developed the MRI-compatible servomotor, oversaw all aspects of the study, and assisted with experiments. All authors contributed to the revision of the manuscript.

# SUPPLEMENTARY MATERIALS

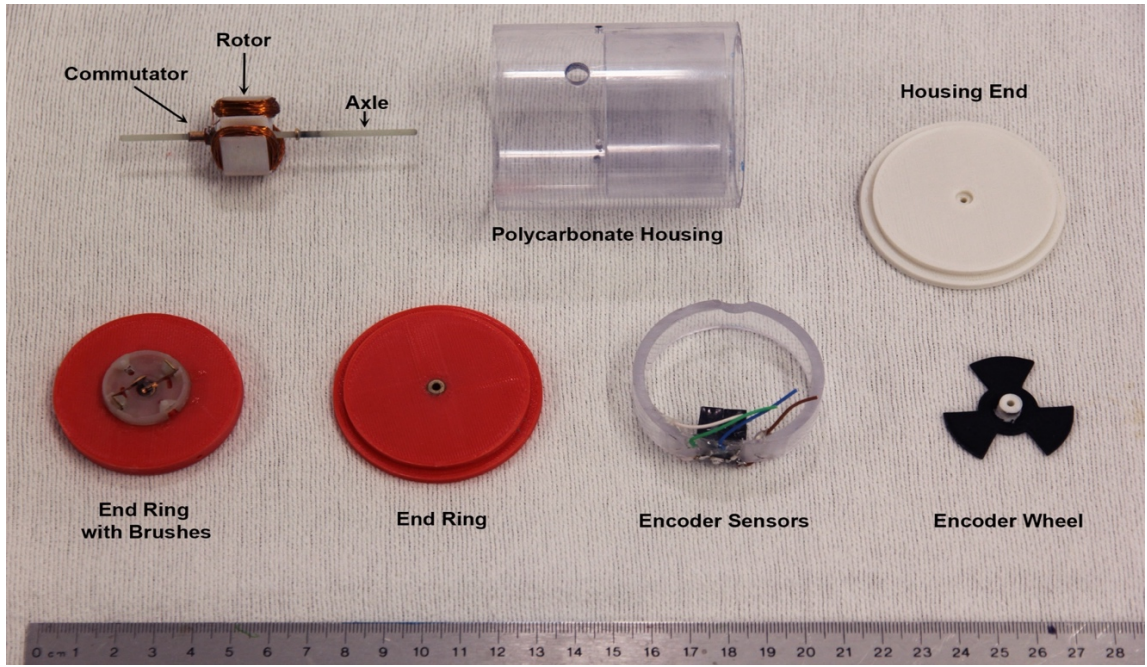

**Fig. S1. Photograph of MRI-compatible DC servomotor components prior to assembly.**

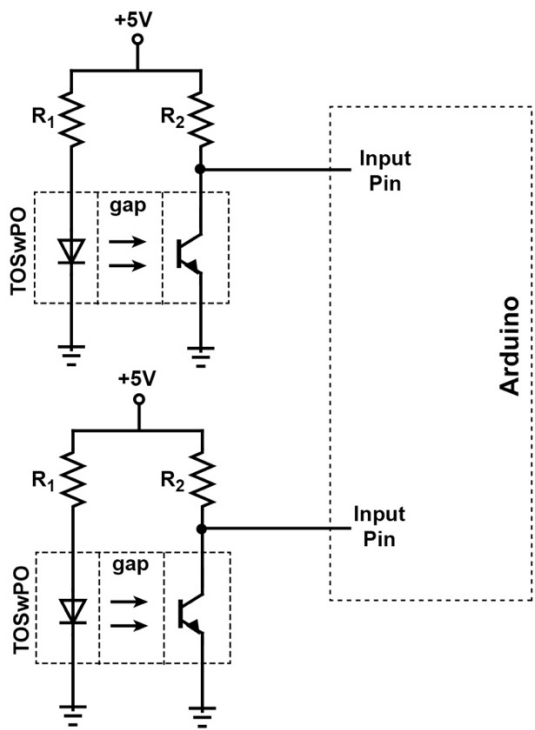

**Fig. S2. Schematic of servomotor optical encoder.** A Cat7 ethernet cable connects the TOSwPO sensors and resistors R1 and R2 where R1 = 180 Ω and R2 = 2 kΩ. Encoder logic is implemented on the Arduino controller.

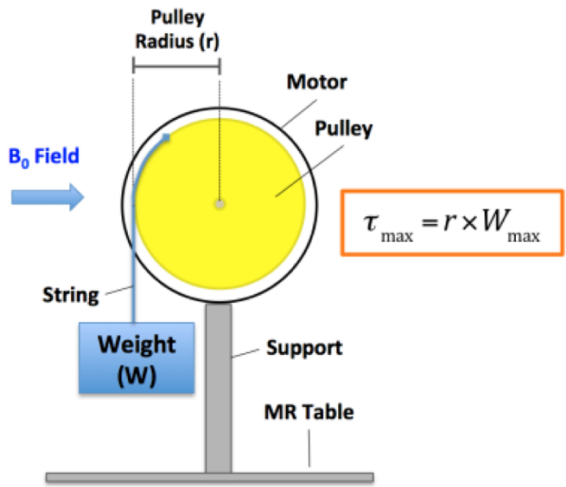

**Fig. S3. Apparatus for measuring stall torque inside MRI.** Stall torque ($\tau_{max}$) was computed by determining the maximum weight ($W_{max}$) that could be lifted at motor stall for a given pulley radius (r).

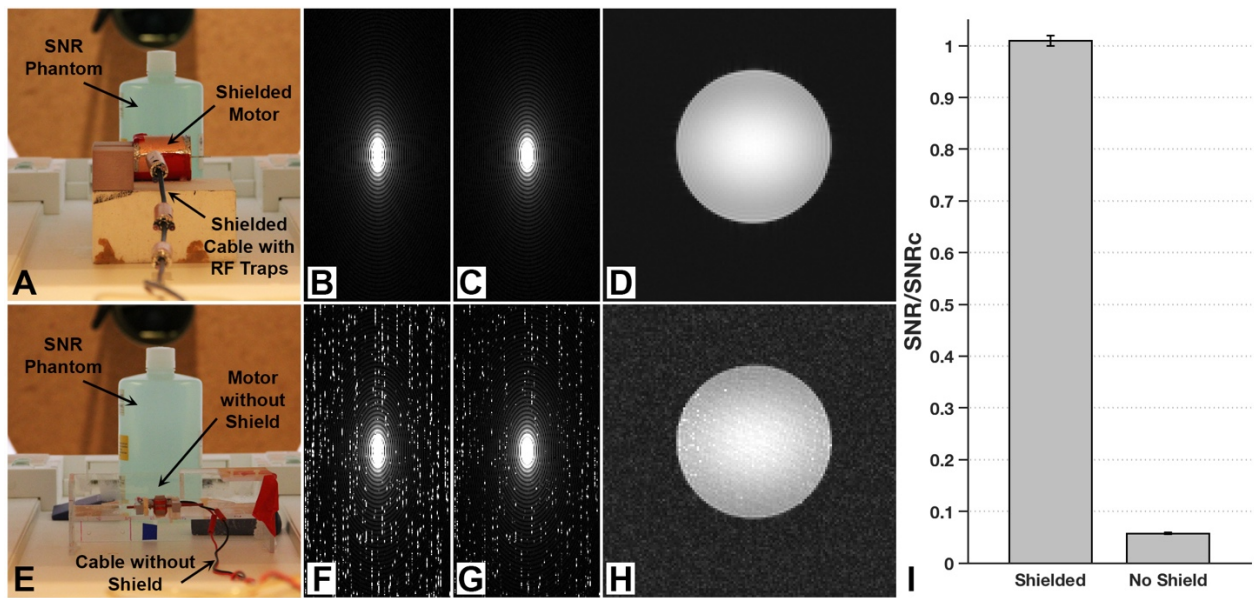

**Fig. S4. Proper EMI shielding and motor design enables simultaneous operation of motor and MRI.** (A) Photograph of the MRI-compatible servomotor located 45 cm from phantom. The raw MRI k-space data acquired during continuous servomotor operation for the two body coil receive channels is shown in (B) and (C). The coil-combined reconstructed magnitude image is shown in (D). (E) Photograph of an unshielded MRI-safe DC motor prototype located 45 cm from phantom. The raw MRI k-space data obtained during continuous DC motor operation for the two body coil receive channels is shown in (F) and (G). The coil-combined reconstructed magnitude image of (F) and (G) is shown in (H). Bright spike artifacts in the raw k-space data in (F) and (G) are readily apparent and image quality in (H) is significantly reduced when compared to (D). (I) Impact on the measured SNR during simultaneous imaging and MRI-compatible servomotor operation was negligible. However, when the unshielded DC motor was operated simultaneous to imaging, the SNR was 5.7% of the control measurement. Data in (I) shows mean +/- SD.

Page 18 of 20

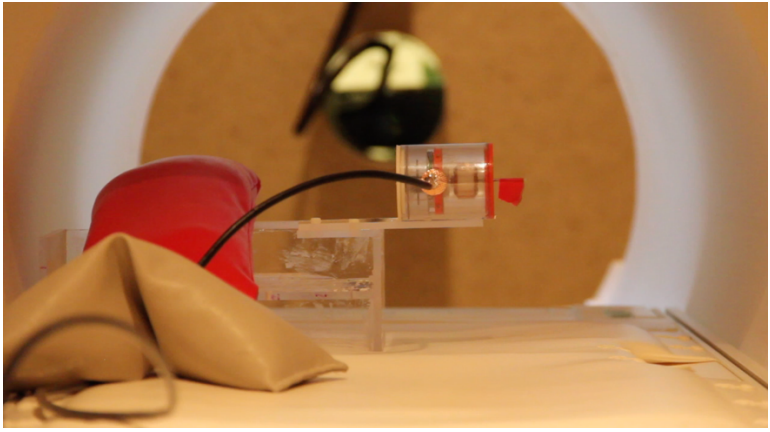

**Video S1.** MRI electromagnetic servomotor without EMI shielding operating in clinical 3T scanner.

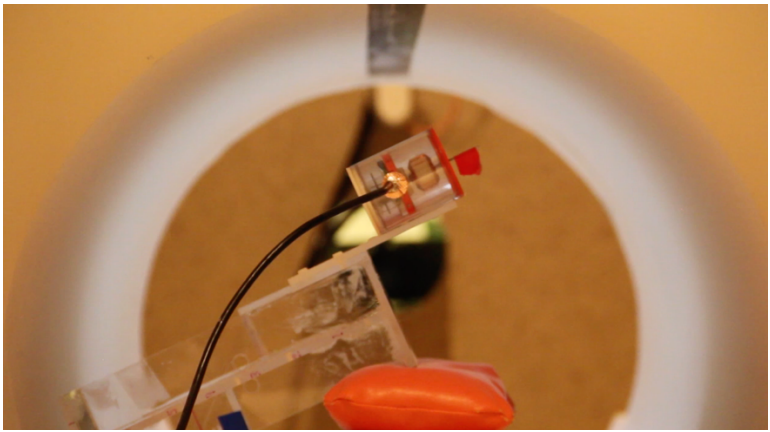

**Video S2.** MRI electromagnetic servomotor without EMI shield operating at an angle in clinical 3T scanner.

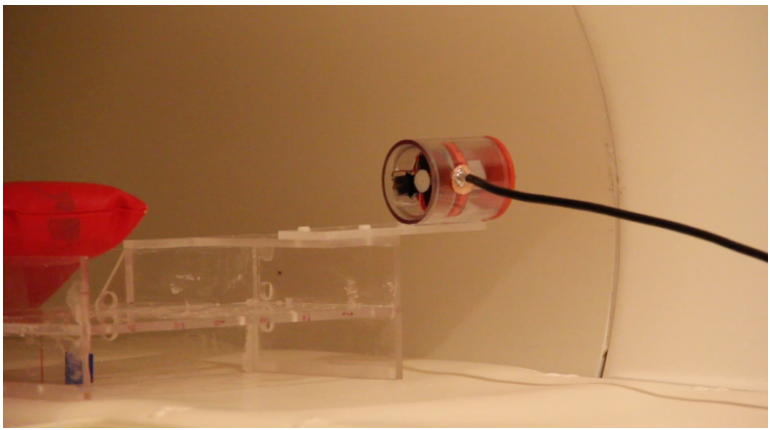

**Video S3.** MRI electromagnetic servomotor maintaining desired setpoint.

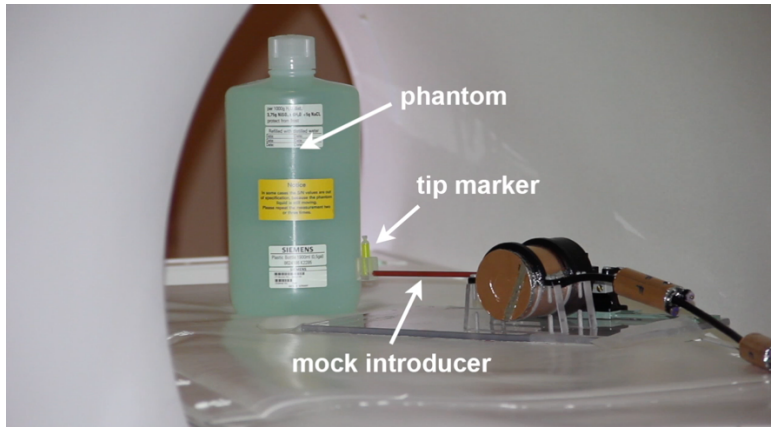

**Video S4.** MRI-compatible biopsy introducer robot with mock introducer demonstrates simultaneous MRI and motor operation when robot oriented at an angle in MR scanner.

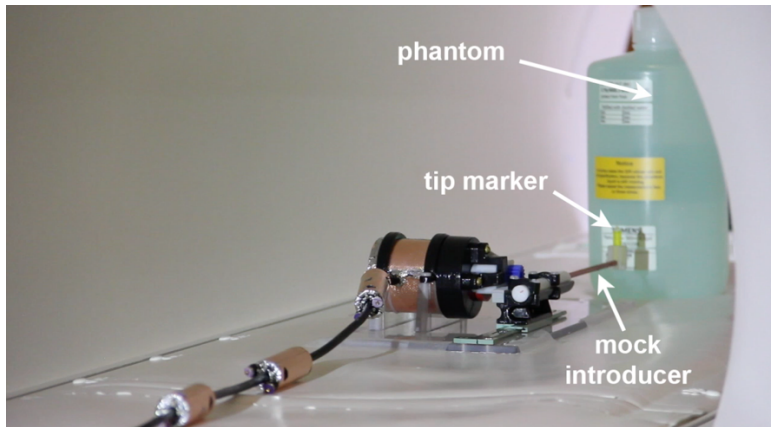

**Video S5.** MRI-compatible biopsy introducer robot with mock introducer demonstrates simultaneous MRI imaging and motor operation.

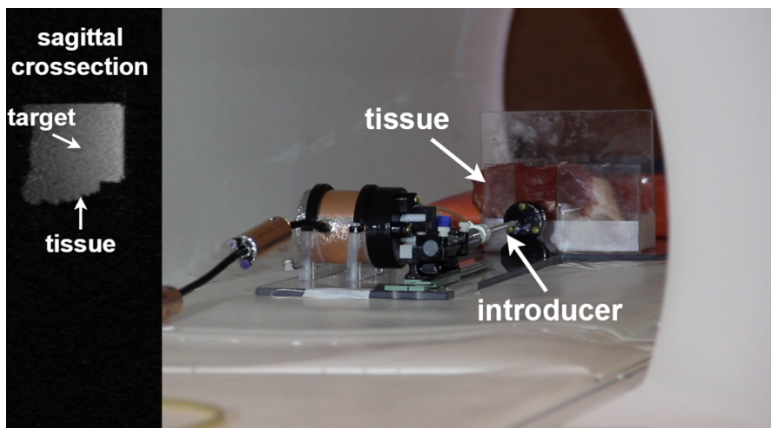

**Video S6.** Placement of 9-gauge introducer sheath using MRI-compatible biopsy introducer robot during continuous imaging.